\documentclass[twocolumn,10pt]{IEEEtran}
\pdfoutput=1 
\usepackage{comment}
\includecomment{toexclude}
\usepackage{amsmath,amsfonts,amssymb}
\usepackage{cite}
\usepackage{verbatim}
\usepackage{bm}
\usepackage{algorithm}
\usepackage[pdftex]{graphicx}
\usepackage{epstopdf}

\newcommand{\yf}{{\mathbf{y}}}

\newcommand{\Yf}{\mathbf{Y}}

\newcommand{\xf}{{\mathbf{x}}}

\newcommand{\A}{\mathbf{A}}
\newcommand{\av}{\mathbf{a}}

\newcommand{\D}{\mathbf{D}}

\newcommand{\B}{\mathbf{B}}
\newcommand{\Pc}{\mathbf{P}}
\newcommand{\p}{\mathbf{p}}

\newcommand{\I}{\mathbf{I}}

\newcommand{\W}{\mathbf{W}}

\newcommand{\diag}{\text{diag}}

\newcommand{\C}{\mathbf{C}}
\newcommand{\cv}{\mathbf{c}}

\newcommand{\U}{\mathbf{U}}
\newcommand{\V}{\mathbf{V}}

\title{Framework for an Innovative Perceptive Mobile Network Using Joint Communication and Sensing}

\author{{J. Andrew Zhang$^1$, Antonio Cantoni$^2$, Xiaojing Huang$^1$, Y. Jay Guo$^1$ and Robert W. Heath Jr$^3$} \\
$^1$University of Technology Sydney, Global Big Data Technologies Centre (GBDTC), Australia\\
$^2$ University of Western Australia, Perth, Australia\thanks{A. Cantoni's work is supported by Discovery Project DP140100522 of the Australian Research Council.} \\
$^3$The University of Texas at Austin, Austin, TX 78712, USA\\
\{Andrew.Zhang; Xiaojing.Huang; Jay.Guo\}@uts.edu.au; antonio.cantoni@uwa.edu.au; rheath@utexas.edu.
}

\begin{document}

\maketitle \thispagestyle{empty} \pagestyle{empty}

\begin{abstract}
 In this paper, we develop a framework for an innovative perceptive mobile (i.e. cellular) network that integrates sensing with communication, and supports new applications widely in transportation, surveillance and environmental sensing. Three types of sensing methods implemented in the base-stations are proposed, using either uplink or downlink multiuser communication signals. The required changes to system hardware and major technical challenges are briefly discussed. We also demonstrate the feasibility of estimating  sensing parameters via developing a compressive sensing based scheme and providing simulation results to validate its effectiveness.
\end{abstract}

%

\section{Introduction}\label{sec:intro}

Radio science and engineering has been advancing in wireless communication and radar sensing in parallel and with limited intersections over a century. Performing joint communication and sensing, however, offers substantial benefits including the potential for shared spectrum and hardware, which is valuable for emerging platforms such as unmanned aerial vehicles and smart cars \cite{Sturm11,Kumari15}. Sensing here refers to information retrieval through measuring  \textit{spatial parameters} such as location and moving speed and \textit{physical parameters} such as shape and hardware characteristics, of static and moving objects using radio signals. Integrating the two functions of communication and sensing into one system can achieve immediate benefits of reduced cost, size, weight, and better spectrum efficiency.  Most existing research however, is limited to point-to-point systems such as millimeter wave radio \cite{Kumari15}, and passive sensing using, e.g.,broadcasting TV signals \cite{Berger10} and WiFi \cite{Chetty12}. 

In this paper, we propose a \textit{perceptive mobile network} using joint communication and sensing techniques,  and identify  basic requirements that need to be met and key issues that need to be resolved. Covering most of the land today, the mobile (i.e., cellular) network is evolving towards a heterogeneous network capable of connecting almost everything. A perceptive mobile network would sense both signal-emitting and silent objects, while providing non-compromised communication services. It can potentially become a ubiquitous sensor, providing seamless radio sensing wherever there is mobile signal coverage, for \textit{detecting, tracking, and identifying objects, activities and events}. Such a radio sensor can facilitate a vast number of new applications that no existing sensor systems can enable, particularly in the various “smartness” initiatives, such as smart city, smart home, smart car and smart transportation. This sensing information can even be used to help establish the communication link \cite{Gonza16}. 


Our work is the first step in developing a basic framework for such a perceptive mobile network, including a description of the structure, feasibility, main challenges, and preliminary sensing algorithms. The limited number of papers that explore mobile signals for sensing, such as \cite{Samcz11, David15}, are constrained to the use of a third device \cite{Samcz11}, and/or involve only simple mobile signals \cite{Samcz11, David15}. In contrast with this work, we investigate a solution seamlessly integrating sensing into the actual signal and network architecture for communications in mobile networks.   

In this paper,  we first explore three types of sensing methods that can be integrated into a mobile network, and the required hardware changes for such an integration. We then formulate the sensing problem and propose a preliminary scheme that can estimate sensing parameters for all the three sensing methods. Numerical results are provided to validate the effectiveness of the proposed framework for this perceptive mobile network. 

\section{Integrated Sensing in Mobile Networks }\label{sec-problem}

We consider a system model with major components aligning with current and 5th generation mobile networks. These components are critical for successful sensing in a mobile network where there could be numerous desired and unwanted echoes or multipath signals. We use the term \textit{clutter} for the unwanted echoes. These system components include:
\begin{itemize}
\item Multiuser multiple-input multiple-output (MIMO). Massive MIMO is preferred but not essential;
\item Multicarrier modulation; 
\item Cloud radio access network (CRAN) architecture: Cooperative remote radio units (RRU) are densely distributed and signal processing for RRUs is done centrally in CRAN central (shortened as base-station, BS.). 
\end{itemize}

A typical communication scenario is as follows: several RRUs work cooperatively to provide connections to mobile stations (MSs), using multiuser MIMO techniques over the same subcarriers. Cooperative RRUs can be within the signal coverage area of each other.
    
\subsection{Three Types of Sensing Methods}
\begin{figure}[t]
 \centering
 \includegraphics[width=0.9\columnwidth]{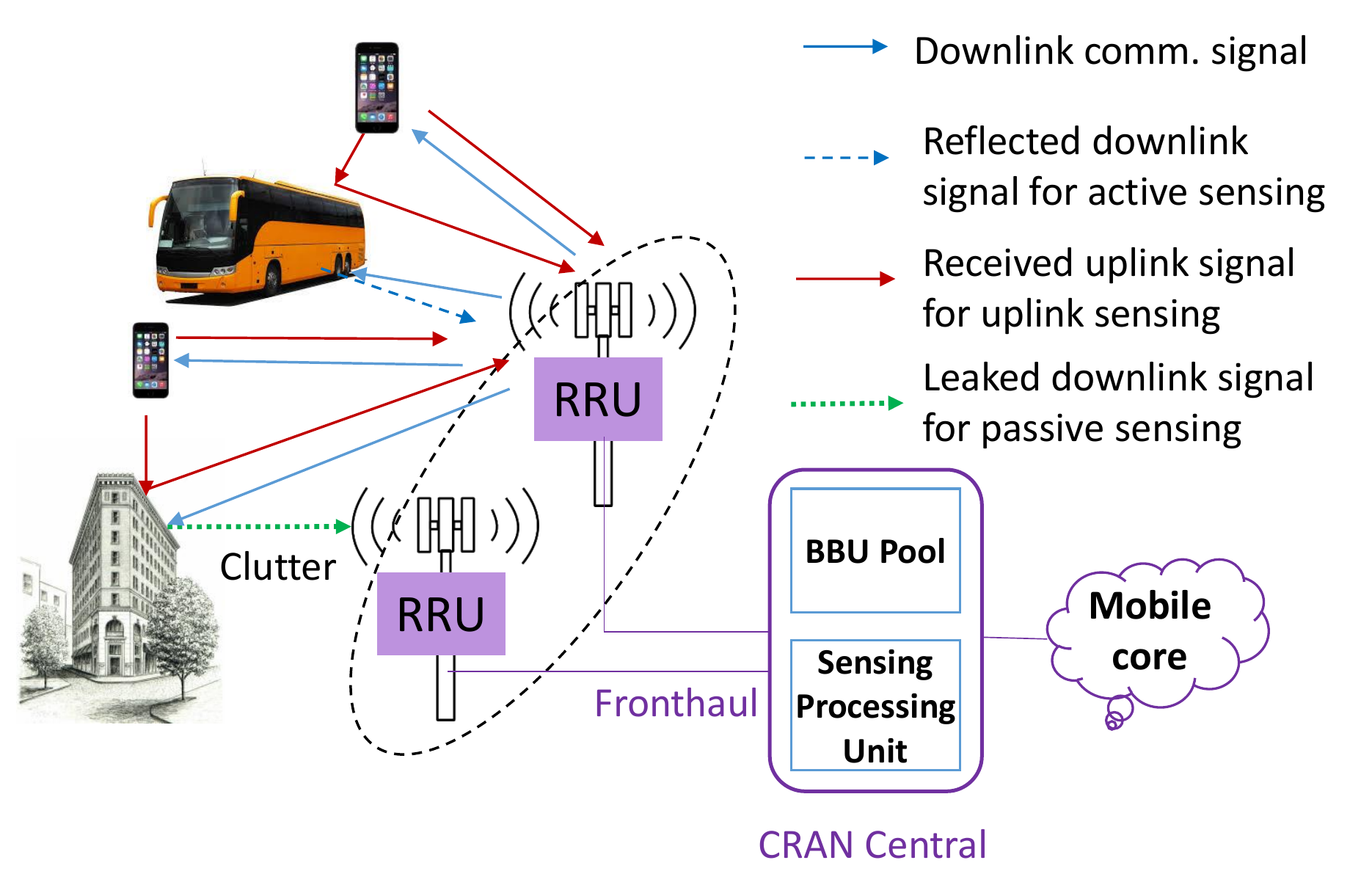}
 \caption{Illustration of three sensing methods in the considered system setup.}
 \label{fig-system}
 \end{figure}

We consider sensing in BSs at the network-side. Three types of sensing methods fit within our framework: \textit{Active Sensing, Passive Sensing} and \textit{Uplink Sensing}. Fig.\ref{fig-system} illustrates the three methods in the considered system setup. Note that RRUs only collect signals, and actual sensing processing is done centrally in BSs.
 
\subsubsection{Active Sensing}
We refer to \textit{active sensing} as the case where a RRU uses reflected signals from its own transmitted signal for sensing. Referring to communication, this is actually the reflected downlink communication signals. In this case, similar to a monostatic radar, transmitter and receiver are co-located although they may have two independent antennas slightly separated in space. This will enable a RRU to sense its surrounding environment. 

\subsubsection{Passive Sensing}

In the case of \textit{Passive sensing} an external third receiver exploits the communication signal for sensing. In our system a RRU uses signals received from other RRUs for sensing. Referring to communication, these will also be the downlink communication signals, and hence active and passive sensing operate at the same stage. Passive sensing senses the environment among RRUs.

We will also refer to active or passive sensing as \textit{Downlink Sensing} when there is no need to differentiate them. One advantage of downlink sensing is that the entire waveform in the transmitted signal and thus received signal is centrally known.  

\subsubsection{Uplink Sensing}
BS uses the uplink communication signal from MS transmitters for \textit{uplink sensing}. Uplink sensing estimates relative, instead of absolute, time delay parameters because the timing in MS transmitters and RRU receivers is not aligned. This ambiguity due to lack of synchronization can be removed by using, e.g., time difference of arrival measurement and the triangulation techniques in localization. Uplink sensing senses MSs and the environment between MSs and RRUs.
 
\subsection{Transceiver Structure and Operations}\label{sec-structure}

Uplink sensing can be implemented without requiring changes to hardware and system architectures of current mobile systems. Downlink sensing requires changes that depend on the duplexing mode. Despite requirements for hardware change, downlink sensing can potentially lead to better sensing results due to, e.g., fixed locations of RRUs, higher transmit power, larger arrays, and more coherent integration time since the transmitted waveform is centrally available.  

\subsubsection{Time Division Duplexing (TDD)}
\begin{figure}[t]
\centering
\includegraphics[width=0.8\columnwidth]{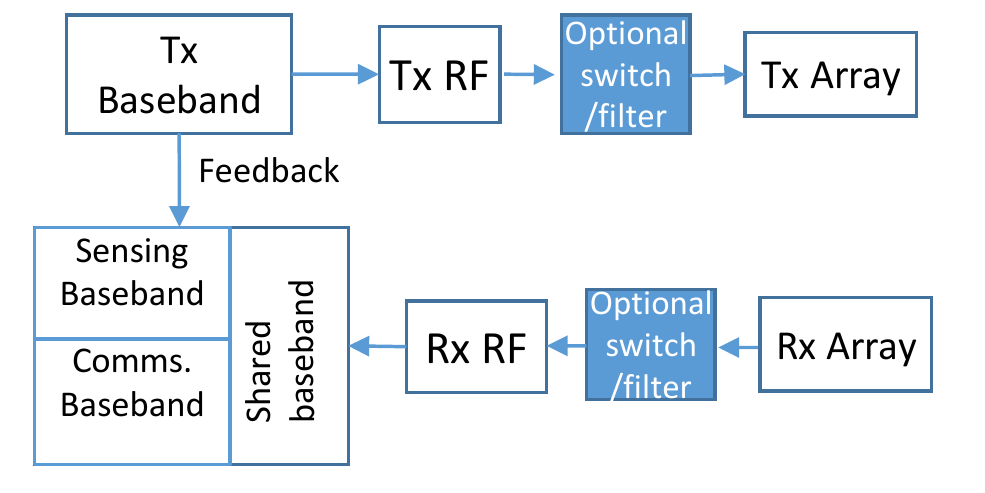}
\caption{System structure of using separated antennas for transmitter and receiver. This is suitable for both TDD and FDD systems.}
\label{fig-trx2}
\end{figure}

A TDD system requires less hardware changes than FDD for downlink sensing. A TDD transceiver generally uses a switch to control the connection of an antenna to the transmitter or receiver. To use downlink signals for sensing, receivers also need to work when transmitter is working. If the switch is always connected to the receiver, the large leakage signal from the local transmitter will bury its own reflected signals and those from remote RRUs and make them non-detectable. Techniques being explored recently for full duplex systems may be applied here, such as RF and digital baseband interference cancellation. These techniques are, however, challenging to implement in real mobile systems. One simpler and viable solution is to separate transmitting and receiving signals by using two sets of separated antennas for transmitter and receiver, as shown in Fig. \ref{fig-trx2}. This requires extra antenna installation space and can increase the overall cost.

During each working cycle in a RRU, the receiver can implement active and passive sensing during the downlink stage, and operate on communication and uplink sensing modes during the uplink stage.

\subsubsection{Frequency Division Duplexing (FDD)}
An FDD transceiver generally uses a diplexer to separate downlink and uplink signals, which are typically well separated in frequency. For downlink sensing, existing hardware in RRUs needs to be changed to enable the receiver to process downlink signals that are in a different frequency band. To this end, more hardware changes to existing systems are required to implement downlink sensing in FDD than in TDD systems.

\subsection{Major Challenges}
Among many challenging problems in developing a perceptive mobile network, clutter suppression and sensing parameter extraction from complex mobile signals are two essential ones to be addressed.

BSs in a mobile network typically see many unwanted echoes that need to be suppressed before or after sensing parameter estimation. Various clutter suppression techniques have been studied in Radar. 
However,  adaptation of these techniques to mobile sensing is not straightforward because the signals and working environment for the two systems are very different. 

Modern mobile signals, particularly in the uplink, could be fragmented  - discontinuous in and varying over time, frequency or space, due to random multiuser access and diverse resource allocation. Extraction of sensing parameters from such complicated signals  is quite challenging.  Most of existing work is not directly applicable here. For example, typical channel estimation algorithms in communications only estimate composited channels with limited unknown parameters \cite{Bajwa10}, and radar systems \cite{Hadi2015} generally use optimized or unmodulated transmitted signals.

\section{Formulation of Sensing Parameter Estimation}
We focus on estimating spatial parameters including distance, direction, and speed of objects in this paper.   

\subsection{System Model}

We consider a CRAN system with $Q$ RRUs and each RRU has $M_q$ antennas configured in the form of a uniform linear array. These RRUs cooperate and provide multiuser MIMO service to $K$ users, using MIMO-OFDMA type of signalling schemes. Each user has  an antenna array of $M_k$ elements. We consider the simple case of $M_q=M$ for any $q$ and $M_k=M_T$ for any $k$ in this paper. Extension of the results to different values of $M_q$ and $M_k$ shall be straightforward. For both uplink and downlink, we assume that data symbols are first spatially precoded, and an IFFT is then applied to each spatial stream. The time domain signals are then assigned to the corresponding RRUs. Let $N$ denote the number of subcarriers and $B$ the total bandwidth. Then the subcarrier interval is $f_0=B/N$ and OFDM symbol period is $T_s=N/B+T_p$ where $T_p$ is the period of cyclic prefix.

Assume a planar wave-front in signal propagation. The array response vector of a size-$M$ array for narrowband signals is given by
\begin{align}
\av(M,\theta)=[1,e^{j\kappa \sin(\theta)},\cdots, e^{j\kappa (M-1)\sin(\theta)}]^T,
\end{align}
where $\kappa=2\pi d/\lambda$, $d$ is antenna interval in the array, $\lambda$ is the wavelength, and $\theta$ is either angle-of-depart (AoD) or angle-of-arrival (AoA). 

For the $\ell$-th out of a total of $L$ multipath signals, let $\theta_{\ell}$ and $\phi_{\ell}$ be the AoD and AoA, respectively, and $b_{\ell}$ the amplitude,  $\tau_{\ell}$ the propagation delay, and $f_{D,\ell}$ the associated Doppler frequency. The basic task for sensing is to estimate these spatial parameters $\{\tau_{\ell},f_{D,\ell},\phi_{\ell}, \theta_{\ell}, b_{\ell}\}, \ell=1,\cdots,L$ from the received signal. In formulating the sensing problem next, we will ignore system imperfections such as carrier frequency offsets and timing offsets between different RRUs, and between MSs and RRUs. 


\subsection{Formulation for Active and Passive Sensing}

According to the system structure in Section \ref{sec-structure}, for any RRU, it sees reflected signals from itself and the multipath signals from the other $Q-1$ RRUs. Its received signal at the $n$-th subcarrier and the $t$-th OFDM block can be represented as
\begin{align}
\yf_{n,t}=&\sum_{q=1}^Q\sum_{\ell=1}^{L_q} b_{q,\ell} e^{-j2\pi n \tau_{q,\ell}f_0}e^{j2\pi t f_{D,q,\ell} T_s}\cdot\notag\\
\label{eq-yf1}
& \quad\av(M,\phi_{q,\ell})\av^T(M,\theta_{q,\ell})\xf_{q,n,t}  +\mathbf{z}_{n,t},\\
=&\A(M,\bm{\phi})\C_n\D_{t}\U^T\xf_{n,t}+\mathbf{z}_{n,t},
\label{eq-active}
\end{align}
where variables with subscript $q$ are for the $q$-th RRU, $\xf_{q,n,t}$ are the transmitted signals at subcarrier $n$ from the $q$-th RRU, 
\begin{align}
&{\A(M,\bm{\phi})}={(\A_1(M,\bm{\phi}_1),\cdots,\A_Q(M,\bm{\phi}_Q))},\\
&{\xf_{n,t}}={(\xf_{1,n,t},\cdots,\xf_{Q,n,t})^T},\\
&\U=\text{diag}\{\A_1(M,\bm{\theta}_1), \A_2(M,\bm{\theta}_2),\cdots,\A_Q(M,\bm{\theta}_Q)\}, 
\label{eq-U2}
\end{align}
and hence $\U$ is an $MQ\times L, L=\sum_{\ell=1}^Q L_q$, block diagonal matrix. The $\ell$-th column in $\A_q(M,\bm{\phi}_q)$ (or $\A_q(M,\bm{\theta}_q)$) is $\av(M, \phi_{q,\ell})$ (or $\av(M,\theta_{q,\ell})$), $\D_t$ and $\C_n$ are diagonal matrices with the $\ell$-th diagonal element being $b_\ell e^{j2\pi tf_{D,\ell} T_s}$ and $ e^{-j2\pi n \tau_{\ell}f_0}$, respectively, $\mathbf{z}_{n,t}$ is the noise vector. Note that multipath signals in the above expression are indexed in the order of RRUs from $1$ to $Q$. 

\subsection{Formulation for Uplink Sensing}\label{sec-direct}

The received signal in a RRU at the $n$-th subcarrier and the $t$-th OFDM block can be represented as
\begin{align}
\yf_{n,t}&=\sum_{k=1}^K\sum_{\ell=1}^{L_q} b_{k,\ell} e^{-j2\pi n \tau_{k,\ell}f_0}e^{j2\pi t f_{D,k,\ell} T_s}\cdot\notag\\
\label{eq-yfnt0}
& \quad\av(M,\phi_{k,\ell})\av^T(M_T,\theta_{k,\ell}) \xf_{k,n,t}  +\mathbf{z}_{n,t},
\end{align}
where symbols have similar meaning and expressions to those in (\ref{eq-active}), except that here they are for MSs instead of RRUs.

Comparing (\ref{eq-yfnt0}) with (\ref{eq-yf1}), we can see that they have similar expressions except for different symbols and parameter values. Hence, next we will develop a general on-grid expression for both downlink and uplink sensing. 

\subsection{Generalized Delay-Quantized On-grid Formulation}

In this paper, we will develop 1-D compressive sensing based algorithms for spatial parameter estimation. We assume that the number of subcarriers $N\gg L$ and $N$ is large enough such that the quantization error of $\tau_{\ell}$ is small and the delay estimation can be well approximated as an on-grid estimation problem. Let the delay term $e^{-j2\pi n\tau_{\ell'} f_0}$ be quantized to $e^{-j2\pi n\ell/(gN)}$, where $g$ is a small integer and its value depends on the method used for estimating $\tau_{\ell}$. The minimal resolvable delay is then $1/(gB)$. 

Let $K$ and $M_T$ denote the total number of users/RRUs and number of antennas in each user/RRU, respectively, for either uplink or downlink sensing. We now convert the multipath signal models in (\ref{eq-yfnt0}) and (\ref{eq-yf1}) to a generalized on-grid (delay only) sparse model, by representing it using $N_p\gg L, N_p\le gB$ multipath signals where only $L$ signals are non-zeros. Referring to (\ref{eq-active}), the delay-on-grid model can be represented as 
\begin{align}
\yf_{n,t}=\A(M,\bm{\phi})\C_n\D_{t}\Pc\U^T\xf_{n,t}+\mathbf{z}_{n,t},
\label{eq-yf4}
\end{align}
where $\C_n$ is now redefined as $\C_n=\diag\{1, e^{-j2\pi n/(gN)},\cdots, e^{-j2\pi n (N_p-1)/(gN)}\}$, re-ordered according to the quantized delay values; $\Pc$ is an $N_p\times L$ rectangular permutation matrix that maps the signals from a user/RRU to its multipath signal, and has only one non-zero element of value $1$ in each row; the other symbols have similar expressions with those in (\ref{eq-active}), with elements in $\A(M,\bm{\phi})$ and $\D_{t}$ being reordered according to the delay. More specifically, the columns in $\A(M,\bm{\phi})$ of size $M\times N_p$ and the diagonal elements in $\D_{t}$ of size $N_p\times N_p$ are now re-ordered and tied to the multipath delay values. $\U$ is an $M_TK\times L$ block diagonal radiation pattern matrix for $M_T$ arrays. $\xf_{n,t}$ is the $M_TK\times 1$ symbol vector. For the moment, we allow repeated delay values in $\C_n$ to account for multipath signals with the same quantized delay but different AoAs and/or AoDs.

\section{Estimation of Spatial Parameters}

We now propose a preliminary scheme based on 1-D compressive sensing for estimating the spatial parameters. This scheme works for all three sensing methods. We assume that the symbols $\xf_n$s are known. For uplink sensing, this can be achieved by demodulating the symbols as sensing can tolerate more delay than communication, while for downlink sensing, they are centrally known. However, the range of subcarriers in downlink and uplink sensing could be different. RRUs can see signals at more subcarriers in downlink sensing than uplink because the total subcarriers could be shared by different group of users. 

\subsection{Single Multipath for Each Delay}

Rewrite (\ref{eq-yf4}) as
\begin{align}
\yf_{n,t}^T 
&=\xf^T_{n,t}(\cv^T_n\otimes\I_{M_TK}) \V\A^T(M,\bm{\phi}),
\end{align} 
where $\otimes$ denotes the Kronecker product, $\cv_n=(1, e^{-j2\pi n/(gN)},\cdots, e^{-j2\pi (N_p-1)/(gN)})^T$, $\I_{M_TK}$ is an $M_TK\times M_TK$ identity matrix, and $\V$ is a $M_TKN_p\times N_p$ block diagonal matrix
\begin{align}
\V=\text{diag}\{b_\ell e^{-j2\pi t f_{D,\ell}T_s}\U\p_\ell\}_{\ell=1,\cdots,N_p},
\label{eq-blockv}
\end{align}
with $\p_\ell$ being the $\ell$-th column of $\Pc^T$. 

We have now separated signals $\xf^T_{n,t}(\cv^T_n\otimes\I_{M_TK})$ that are known and dependent on $n$ from others. Let $\mathcal{S}_s$ denote the set of available subcarriers for sensing and let $N_s$ denote its size. Stacking all row vectors $\yf_{n,t}^T, n\in\mathcal{S}_s$ to a matrix generates
\begin{align}
\Yf_t 
= \W\V\A^T(M,\bm{\phi}),
\label{eq-vat}
\end{align}
where $\W$ is an $N_s\times M_TKN_p$ matrix with its $n$-th row being $\xf^T_{n,t}(\cv^T_n\otimes\I_{M_TK})$.

Inspecting (\ref{eq-vat}), we can see that the estimation problem can be formulated as a multi-measurement vector (MMV) block sparse problem \cite{liu2013energy} with $N_s\times M$ observations $\Yf_t$, sensing matrix $\W$, and block sparse signals $\V\A^T(M,\bm{\phi})$ of $L$-sparsity. Let $\V=(\V_1^T,\V_2^T,\cdots,\V_{N_p}^T)^T$ where $\V_\ell$ denotes the $M_TK\times N_p$ block signals, and $L$ out of $N_p$ $\V_\ell$s have non-zero elements. A block sparse compressive sensing algorithm can solve (\ref{eq-vat}) and generate estimates for $\V_\ell A^T(M,\bm{\phi}), \ell=1, \cdots, N_p$.

We first consider noiseless cases. Once the $L$ nonzero blocks $\V_\ell\A^T(M,\bm{\phi})$ are determined, we can then get the $L$ delay estimates according to the indexes of the blocks.
 
From (\ref{eq-blockv}) we can see that only the $\ell$-th column in $\V_\ell$ has non-zero elements $b_\ell e^{-j2\pi t f_{D,\ell}T_s}\U\p_\ell$ if $b_\ell\neq 0$. Therefore, 
\begin{align}
\V_\ell\A^T(M,\bm{\phi})=b_\ell e^{-j2\pi t f_{D,\ell}T_s}\U\p_\ell\av^T(M,\phi_\ell).
\end{align}
Since $\p_\ell$ only has a single non-zero element $1$, $\U\p_\ell$ will generate a column vector corresponding to one column in $\U$. Because $\U$ is a block diagonal matrix, only $1$ out of $K$ $M_T\times 1$ vectors in each column is non-zero. 

Now represent $\V_\ell\A^T(M,\bm{\phi})$ as $K$ $M_T\times M$ sub-matrices $(\B_{\ell,1}^T,\cdots,\B_{\ell,K}^T)^T$. If $\B_{\ell,k}\neq 0$, then this multipath is from the $k$-th RRU (user).  We can also see that 
\begin{align}
\B_{\ell,k}=b_\ell e^{-j2\pi t f_{D,\ell}T_s}\av(M_T,\theta_{k,\ell})\av^T(M,\phi_{k,\ell}).
\end{align}
From $\B_{\ell,k}$, calculating the cross-correlation between columns and rows, we can obtain AoA or AoD estimates, depending on the order of the calculation.

Doppler shift $f_{D,\ell}$ can be estimated across multiple OFDM blocks, based on the cross-correlation of $\B_{\ell,k}$ in these blocks.

The absolute value of $b_\ell$ can be estimated as the mean power of all elements in $\B_{\ell,k}$. A better estimate is to use the cross-correlation output for estimating AoA. 

In noisy cases, we can sort the blocks $\V_\ell\A^T(M,\bm{\phi}), \ell=1,\cdots,N_p$ according to the estimates of $b_\ell$ and use a threshold to filter out blocks corresponding to multipath signals. We can also keep the estimated results for a subset of $N_p$ blocks with larger estimated $b_\ell$s, and then apply data fusion techniques over all measurements spanning a segment of space, time and frequency domains to get synthesized sensing results. 
 


\subsection{Multiple Multipath Signals with the Same Delay}
Let $\cv_n=(\cv_{n,1}^T,\cv_{n,2}^T,\cv_{n,2}^T)^T$, where $\cv_{n,2}$ represents the repeated entries. We can accordingly represent $\W=(\W_1,\W_2,\W_2)$ and $\V=(\V_1^T,\V_2^T,\V_3^T)^T$. Then
\begin{align*}
\W\V\A^T(M,\bm{\phi})=(\W_1,\W_2)\left(\begin{array}{cc}
\V_1\A^T(M,\bm{\phi})\\
(\V_2+\V_3)\A^T(M,\bm{\phi})
\end{array}\right).
\end{align*}  
This shows that we can always use a $\cv_n$ with single entry for each quantized delay, and multiple signals with different angles will show up in the MMV estimates. More specifically, if $\ell\in \mathcal{S}_d$ multipath signals have the same delays but different AoAs or AoDs, we will get
\begin{align}
\V_\ell\A^T(M,\bm{\phi})=\sum_{\ell\in \mathcal{S}_d}b_\ell e^{-j2\pi t f_{D,\ell}T_s}\U\p_\ell\av^T(M,\phi_\ell).
\end{align}

If these multipath signals are from different RRUs (users), multiple $\B_k$s will be non-zero. Hence in this case, these multipath signals can be estimated straightforwardly.

If multipaths are from the same RRU (user), we will have
\begin{align}
\B_{\ell,k}=\sum_{\ell\in \mathcal{S}_d} b_\ell e^{-j2\pi t f_{D,\ell}T_s}\av(M_T,\theta_{k,\ell})\av^T(M,\phi_{k,\ell}).
\label{eq-Bq2}
\end{align}
When the multipath number is small, we can solve this problem by applying spectrum analysis techniques, such as ESPRIT or MUSIC. We can also right-multiply $M\times 1$ beam-scanning vectors to $\B_{\ell,k}$ to get the magnitude plot at different AoAs.
 
\section{Simulation Results}\label{sec-simu}

We present preliminary simulation results here using the block Bayesian Sparse Learning algorithm \cite{liu2013energy} to validate the effectiveness of our parameter estimation scheme, with $g=1$. 
 
We consider a system with $4$ RRUs, providing connections to $4$ users using all $N=1024$ subcarriers and $4\times 4$ multiuser MIMO. Each RRU has 4 antennas at an interval of $d=\lambda/2$ and each MS has 1 antenna. The carrier frequency is $2.35$ GHz and the signal bandwidth is $100$ MHz. No radar cross-section information is considered. We use a free-space pathloss model but with a pathloss factor of 3, as it is typically in the range of 2 to 4 for the three types of sensing. The total thermal noise in the receiver is $-97$ dBm. Multipath signals for each RRU/MS are generated randomly in a cluster, mimicking reflected/scattered signals from objects. Each cluster is generated following uniform distributions of $[10, 15]$ for the total multipath number, $[0,45]$ degrees for direction span, $[0, 90]$ m for distance, and $[0, 600]$ Hz for Doppler frequency.  Across clusters there are additional offsets in direction, distance and Doppler frequency. Delays are on grid with an interval of 10 ns, corresponding to a distance resolution of $3$ m. Delays from the same RRU/MS are kept different. But they could be the same between RRUs/MSs. Random continuous values are used for Doppler shift, AoAs and AoDs. 

\begin{figure}[t]
\centering
\includegraphics[width=1\columnwidth]{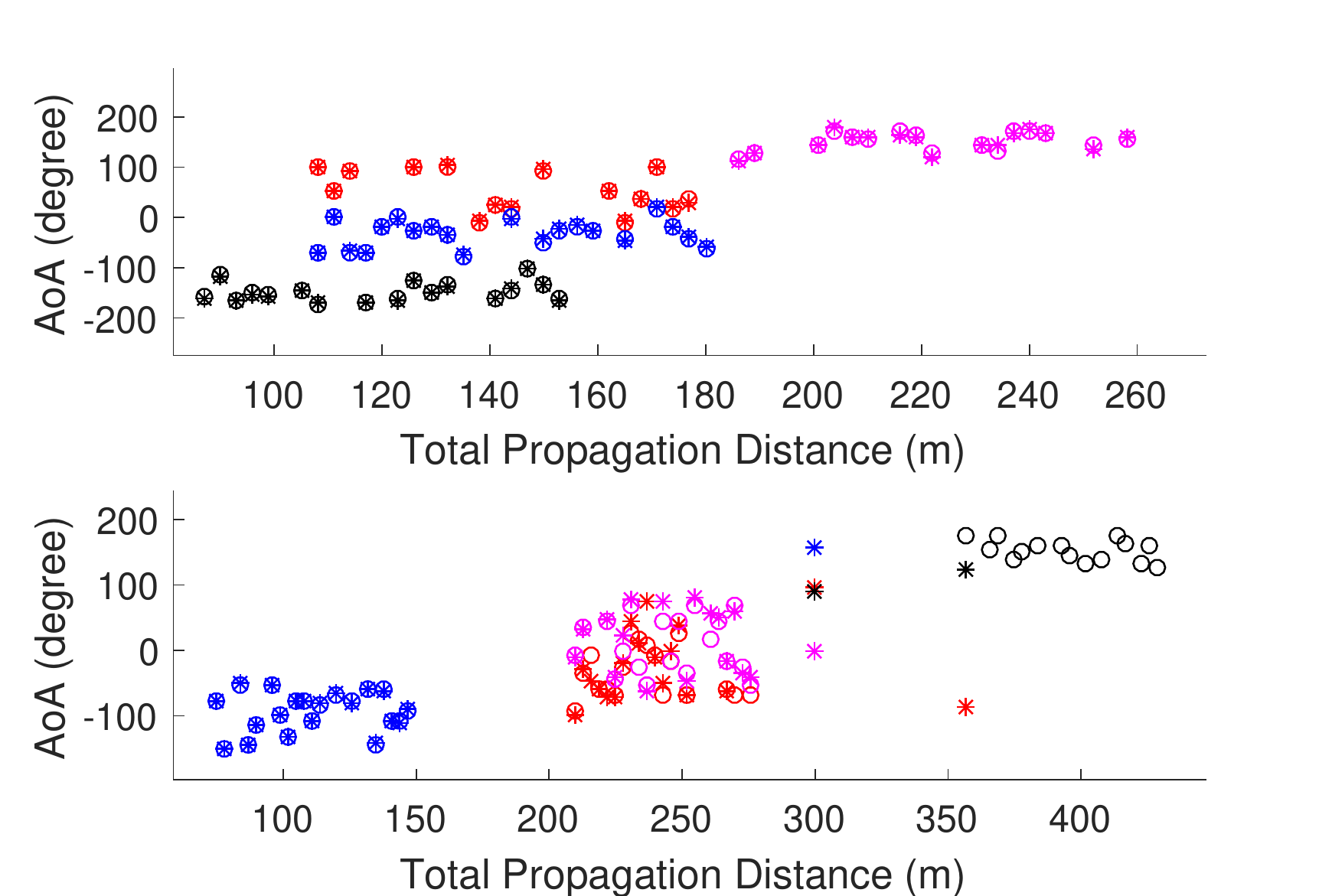}
\caption{Results for AoA-Distance estimation in downlink sensing. Note the AoA here and in Fig. \ref{fig-up} is actually the angle of $e^{j\pi\sin(\phi_{\ell})}$. Every star or circle represents parameters for one multipath: Stars and circles are for estimated and actual ones, respectively. Different colors represent multipath from different RRUs. Upper: transmission power=20 dBm; bottom: transmission power=15 dBm. }
\label{fig-down}
\end{figure}

\begin{figure}[t]
\centering
\includegraphics[width=1\columnwidth]{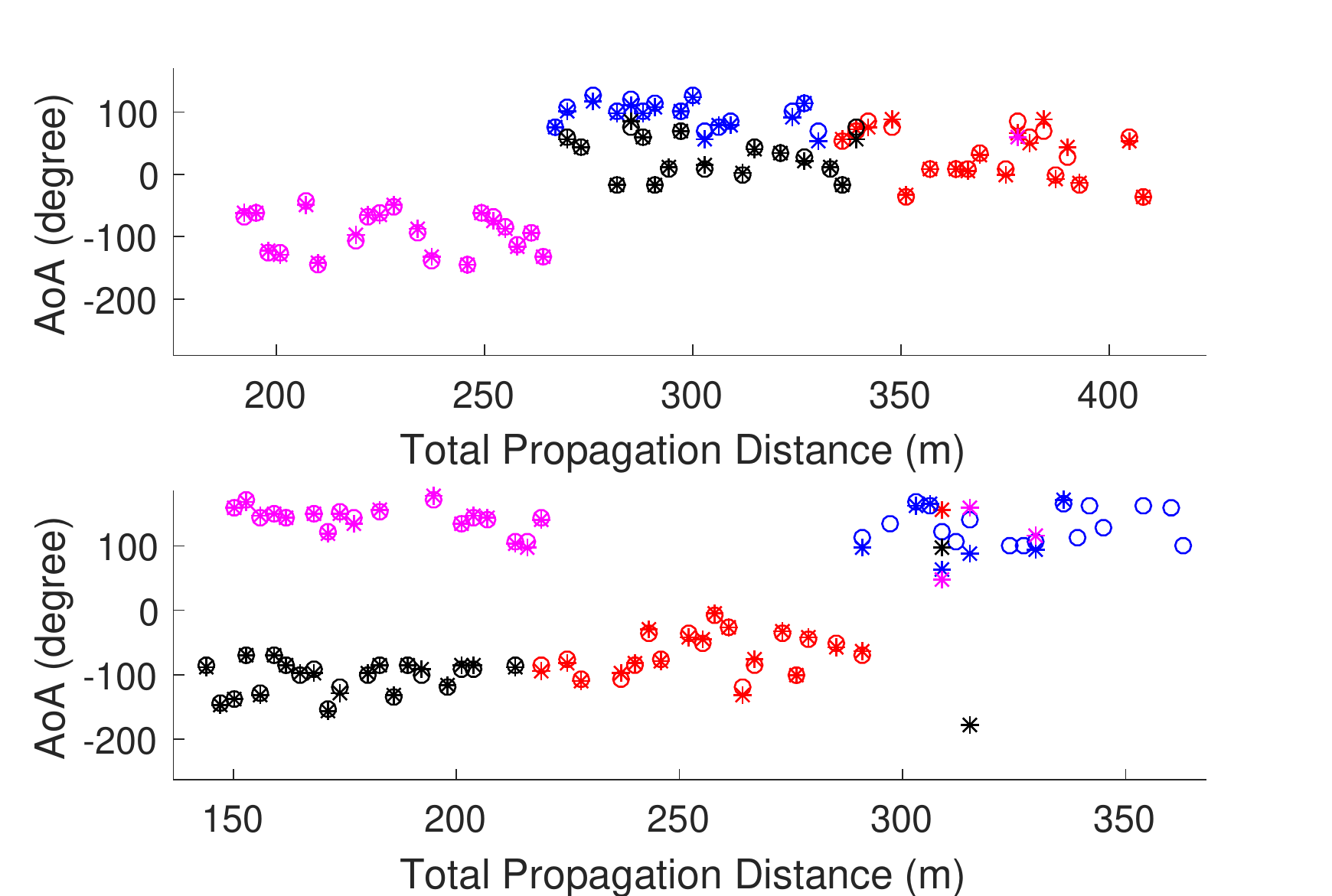}
\caption{Results for AoA-Distance estimation for uplink sensing. Notes are similar to those in Fig. \ref{fig-down}. }
\label{fig-up}
\end{figure}

Figs. \ref{fig-down} and \ref{fig-up} present typical AoA-Distance results for downlink and uplink sensing respectively. The depicted distance is the total signal travelling distance between a transmitter and the receiver, and does not necessarily translate to the distance of objects to the receiver directly. Complex across-RRU synthesizing is needed to achieve this translation, particularly for uplink sensing. For the downlink, all subcarriers are used, and for uplink, 256 interleaved subcarriers are used. Both figures demonstrate that when signal-to-noise ratio (SNR) is sufficiently high, the estimated results show great match with the actual ones. When the SNR decreases, estimation accuracy for multipath signals with lower power degrades. 

\section{Conclusions}
We have developed a basic framework for a perceptive mobile network where three types of sensing methods can be integrated with communication. A preliminary scheme is developed for estimating sensing parameters, and its effectiveness is validated by simulation results. Although there are significant challenges in making the system fully functional, our work here is a solid first step, demonstrating the feasibility and providing a way to proceed.

\bibliographystyle{IEEEtran}
\bibliography{jian}  

\end{document}